# Polymer/paper-based double touch mode capacitive pressure sensing element for wireless control of robotic arm

Rishabh B. Mishra[1, 2], Wedyan Babatain[2], Nazek El-Atab[2], Aftab M. Hussain[1], Muhammad M. Hussain[2, 3, *]
[1]PATRIoT Lab, CVEST, International Institute of Information Technology (IIIT), Hyderabad, Telangana, 500032, India
[2]MMH Labs, Electrical Engineering, King Abdullah University of Science and Technology (KAUST), Thuwal, 23955-6900, Saudi Arabia
[3]Electrical Engineering and Computer Sciences (EECS), University of California, Berkeley, CA, 94720, United States of America (USA)
[*]Email – muhammadmustafa.hussain@kaust.edu.sa or mmhussain@berkely.edu

*Abstract*—In this work, a large area, low cost and flexible polymer/paper-based double touch mode capacitive pressure sensor is demonstrated. Garage fabrication processes are used which only require cutting, taping and assembly of aluminum (Al) coated polyimide (PI) foil, PI tape and double-sided scotch tape. The presented pressure sensor operates in different pressure regions i.e. normal (0 to 7.5 kPa), transition (7.5 to 14.24 kPa), linear (14.24 to 54.9 kPa) and saturation (above 54.9 kPa). The advantages of the demonstrated double touch mode capacitive pressure sensors are low temperature drift, long linear range, high pressure sensitivity, precise pressure measurement and large die area. The linear output along with a high sensitivity range (14.24 to 54.9 kPa pressure range) of the sensor are utilized to wirelessly control the movement of a robotic arm with precise rotation and tilt movement capabilities.

## I. INTRODUCTION

Silicon based and solid-state materials have played a major role in the development of micro-electromechanical system (MEMS) based devices [1], which have been adopted from complementary metal oxide semiconductor (CMOS) industries. As a result, the MEMS devices became compatible with fabrication of integrated circuits (ICs) [2-3]. MEMS technology received attention for designing high performance, versatile and reliable devices for automobile, aerospace, environment, industrial, portable and consumer electronics application [4-6]. The advantages offered by CMOS technology were limited to batch fabrication of rigid and bulky devices due to their state-of-the-art nature [5-9]. However, due to the advancement of free-form CMOS technology, fully flexible and stretchable devices came into existence. Flexible and stretchable electronics play a vital role in the advancement of soft robotics, wearable/healthcare and implantable devices considering the internet of things (IoT), internet of everything (IoE) and embedded electronics applications [7-14].

Paper, polymer and do-it-yourself (DIY) electronics have been utilized to design low-cost systems and devices for flexible electronics applications such as health monitoring, plant monitoring and environmental monitoring [16-20]. Using paper as part of the sensory element has received significant attention due to its various features such as low cost, environmental friendliness, portability, flexibility and low weight which require only few garage fabrication steps like cutting, folding, bending, sculpting, photo-lithography and printing to design large-area flexible electronic systems [20-24]. Fully flexible systems fabricated using paper-based materials have been previously demonstrated for several applications such as health monitoring of body vitals like body temperature, abnormal sweating, heartrate, blood-pressure and wheezing [17-18]. Paper-based electronic devices like smart toy, loudspeaker, cube lamp, flowerpot and architectural model have been designed using paper sculpting methods [21]. Paper substrate, on which polylactide/graphene composite conductive filament is printed, has also been used for actuation and sensing from which modular robot, mimosa leaves and lampshade have been introduced [22]. Similarly, paper-based cantilevers along with carbon and silver inks have been used to design paper-based MEMS piezo-resistive pressure sensor and weighing machine as well [23]. Pencil drawn structures on paper cantilever have been utilized to design the strain sensor and chemiresistor which changes its resistance due to variations in the chemical environment in which the sensor is placed [24].

Among MEMS pressure sensor's arena, capacitive pressure sensors have several advantages like low temperature drift, high sensitivity, simple structure, low power consumption and lower sensitivity to stress at the edges of the mechanical sensitive element which will deform/deflect after pressure application [5-6, 25-26]. However, the nonlinearity between capacitance change versus applied pressure and effect of stray capacitance are major issues when the pressure sensor is operating in the normal mode [21-22]. In the normal mode of capacitive pressure sensor, the deflection in mechanically sensitive diaphragm needs to be less than one third of the separation gap between parallel plate capacitive pressure sensors. The touch-mode capacitive pressure sensor has been presented to overcome the non-linearity and stray capacitance limitations. In this design, the top diaphragm comes into contact with the fixed bottom plate due to pressure application. Touch-mode pressure sensors offer high pressure range, linear response and high overload protection. However, touch mode capacitive pressure sensors have disproportionate relationship between capacitance change v/s applied pressure and a short linear region which causes the sensor to reach saturation mode very fast [25-26, 31]. Thus, continuous efforts have been made to develop approaches aimed at enhancing the linearity region and proportionality of capacitance variation with respect to applied pressure such as double side diaphragm touch mode [32-33], island structure sensor [34], sandwich diaphragm [35] and double touch mode [28, 37] capacitive pressure sensors. However, sealing vacuum cavity between both electrodes has been a major concern during the design of MEMS-based capacitive pressure sensors. If the cavity is not properly sealed, then it may affect the output of sensor which may result in an undesired offset.

This work is supported by the King Abdullah University of Science and Technology (KAUST) Office of Sponsored Research (OSR) under Sensor Innovation Initiative Award No. OSR-2015-Sensors-2707 and KAUST–KFUPM Special Initiative Award No. OSR-2016-KKI-2880.

In paper-based capacitive pressure sensor, microfiber wipe, sponge and air have been reported as dielectric materials between parallel plate electrodes in which the response of sensor with air dielectric has the best response among all [17-18, 20]. To use air as the separation gap, the edges of mechanical sensitive diaphragm have been clamped using double-sided tape and aluminum foil, paper and Al coated PI sheet [18, 20]. Cantilevers are mechanical elements which enhance the sensitivity of capacitive pressure sensor when it comes to measurement of small range of pressure such as acoustic pressure sensing [38]. The normal mode capacitive pressure sensor with different diaphragm shapes has been analyzed for sensitivity, and circular shape capacitive pressure sensor shows highest sensitivity, however, it shows highly non-liner response [25, 39, 40]. The major reason for non-linearity is a large deflection in the diaphragm. Moreover, in normal mode of capacitive pressure sensor, the maximum diaphragm deflection in pressure sensitive diaphragm should be less than 1/3$^{rd}$ of separation gap between parallel plate of capacitor to avoid pull in phenomena. Hence, the solution for avoiding non-linearity and increasing the pressure application range in touch mode capacitive pressure sensor as artificial skin module is presented using low cost paper, Al coated PI and double-sided scotch tape with garage fabrication technique [32].

In this work, double touch mode capacitive pressure sensor has been fabricated and characterized using low cost materials such as Al coated PI sheet, double sided tape and PI tape. Designing the double touch mode capacitive pressure sensor using MEMS technology requires cleanroom fabrication facilities with complex processes [28]. However, we utilized a garage fabrication process to fabricate the sensor which requires just a few steps like cutting the circular shapes, peeling out and placing of tapes which enables mass fabrication of large area pressure sensors. We used a $CO_2$ Laser (Universal Laser Systems PLS6.75) to cut the circular shapes, while a Keithley Semiconductor Characterization System (Model – 4200 SCS) was used to perform capacitance measurements for characterizing the sensor. We used the output of the double touch mode capacitive pressure sensor to wirelessly control the position of a DC servomotor of robotic gripper system which can be utilized for conventional/soft robotic application like gripping.

## II. SENSOR'S DESIGN AND MATHEMATICS

A schematic diagram of a double touch mode capacitive pressure sensor is shown in Fig. 1. It operates in four different regions which are normal, transition, linear and saturation regions. In normal region, the deflection in the mechanically sensitive diaphragm is less than one third of separation gap between plates. As pressure increases, the sensor starts operating in the transition region after the mechanical pull-in phenomenon in which the diaphragm touches the bottom plate. There are two touch points in the transition region. The sensor touches bottom dielectric layer first and the pressure at which this happens is known as "first touch point pressure". As the applied pressure is increased, the sensor touches the upper dielectric layer, known as "second touch point pressure". Beyond this pressure, the sensor operates into near linear region. Finally, it switches to saturation region as a large area of diaphragm already touches the bottom plate.

Deflection in mechanically sensitive element due to pressure application in normal mode MEMS capacitive pressure sensors follows small deflection and Kirchhoff's plate theory. However, for the touch mode capacitive pressure sensors, the deflection in pressure sensitive diaphragm follows large deflection theory.

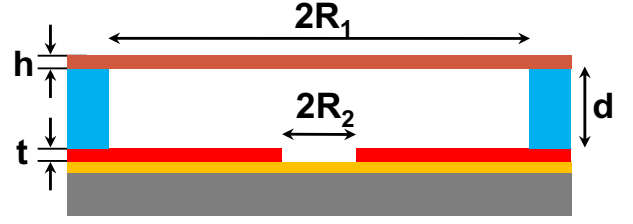

Fig. 1. Schematic of double-touch mode capacitive pressure sensor.

The deflection in mechanically sensitive circular shape diaphragm of capacitive pressure sensor is given by [25, 26, 29, 39]:

$$w(r) = W\left(1 - \left(\frac{r}{R_1}\right)^2\right)^2 \quad (1)$$

where, $R_1$ is radius of diaphragm and W is maximum deflection at the center which can be given by (from the perspective of large deflection theory) [29]:

$$0.488\frac{W^3}{h^2} + W = \frac{PR_1^4}{64\,D} \quad (2)$$

The basic capacitive pressure sensor consists of a parallel plate capacitor in which one plate will be fixed and another one will be mechanically sensitive due to pressure application. The base capacitance of capacitive pressure sensor is given by:

$$C_b = \frac{\epsilon A}{d} \quad (3)$$

where, $\epsilon$, $A$ and $d$ represent permittivity of air, overlapping area and separation gap between plates, respectively.

The diaphragm deflection due to pressure application causes the capacitance change which can be given by:

$$C_P = \epsilon \iint_S \frac{dA}{d - w(r)} \quad (4)$$

where, S is the surface integral of pressure sensitive diaphragm. The mathematical modeling, numerical simulation and fabrication of double touch mode capacitive pressure micro-sensor has been carried out and compared with touch mode capacitive pressure sensor [29, 37]. However, the large overlapping area of the sensor, parasitic effects, and electrical lead transfer for the connection will also affect the response of the sensor [5, 29-30].

## III. FABRICATION FLOW OF SENSOR

The schematics of the fabrication process steps of double touch mode capacitive pressure sensor is shown in Fig. 2. The shape of pressure sensitive diaphragm is chosen to be circular due to its high sensitivity among all different shapes viz. elliptical, square, rectangular, pentagon in MEMS and paper electronics [25, 36, 39-40]. In our approach, the fabrication of the sensor requires only Al coated PI sheet [h = thickness (Al and PI) = 25 μm], PI tape and double-sided tapes. For two pieces of circular shaped sensors [diameter = $2R_1$ = 2.6 cm], the flexible Al coated PI sheet were cut and employed for top and bottom electrodes [Fig. 2(a)]. Al operates as one of the conductive plates for the capacitive pressure sensor. The PI tape with the small concentric circular hole [diameter = 2R$_2$ = 5mm] with the same outer diameter as bottom electrode is placed on the PI side of the bottom electrode [Fig. 2 (b)]. Then, three layers of double sided tape [strip width = 2.5mm] were placed along the edges of the circular shape [Fig 2 (c-d)]. On top of the three layers of double-sided tape, the circular shape of Al coated PI sheet is placed [Fig. 2 (e-f)]. Following these simple steps, the sensor has been fabricated which only requires the use of a laser cutting tool to ensure accurate dimensions of device. The cross sectional view of the sensor is shown in Fig. 2 (g) and the scanning electron microscopic (SEM) images of the sensor at the centre and edges are shown in Fig. 2 (h) and Fig. 2 (i), respectively. The thickness (*t*) of PI tape (which has the small circular hole of 2R$_2$ diameter) is 15 μm. The achieved separation gap between PI tape and mechanically sensitive diaphragm is 425 μm.

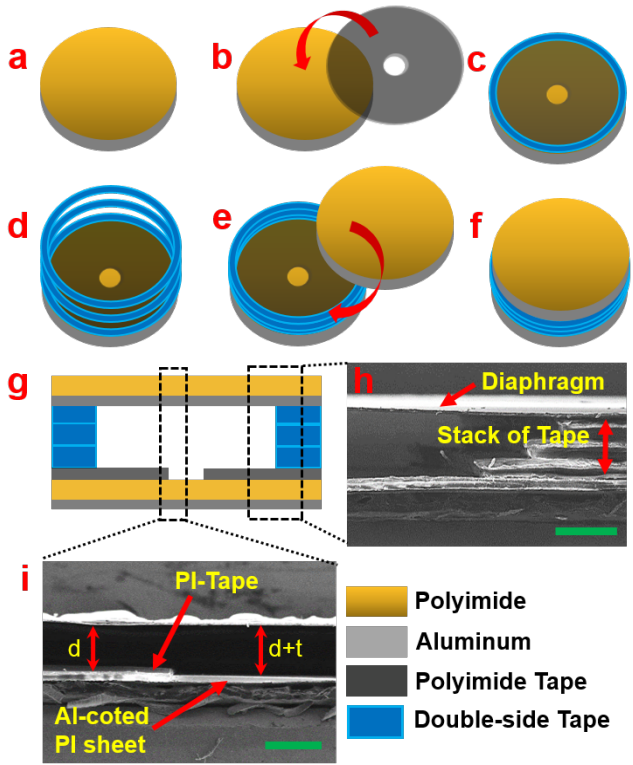

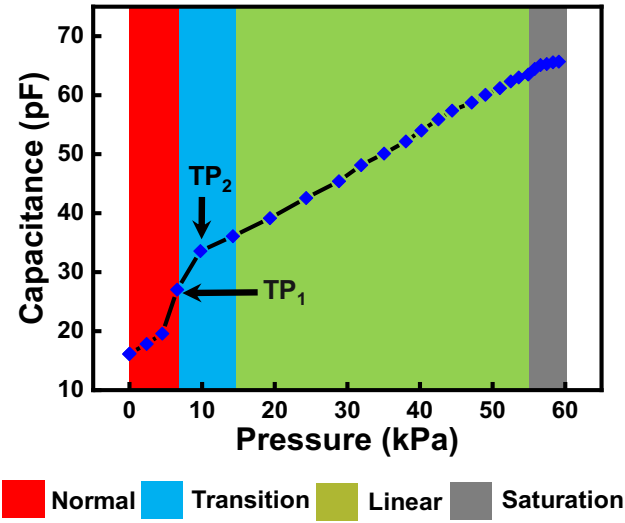

Fig. 3. Output characterstics of double-touch mode capacitive pressure sensor for the pressure range of 0-60 kPa. The sensor operates at normal, transition, linear and saturation region which are shown by different colour bands i.e. red, sky blue, green and gray, respectivelly. There are two different touch points (TP$_1$ and TP$_2$) which are at 7.5 kPa and 9.7 kpa. TP$_1$ is the touch point when the diaphram touches the bottom cavity which is at distance of d+t and TP$_2$ is the touch point when diapgragm touchs the PI-tape which is at distance of d. The linear range of pressure sensor is for 14.24 kPa to 54.9 kPa in which the change in capacitance is from 36.09 pF to 63.5 pF that means the sensor has the sensitivity of 0.674 fF/Pa.

Fig. 2. Schematic diagram of the fabrication process of sensors. (a) Laser cutting the circular shep from the Al coted PI sheet of 2.6 cm diameter. (b). Placing the PI tape with 2.6 cm of external diameter and the 5 mm diameter of inner circular cut is placed on the PI side of previous circualar shape. (c) Placing the single and layer of double sided at the edges. (d) Repeatig the previous steps for two more times. (e) Placing the circular shape, which can obtained from the first step, on top of the third layer of double sided tape. (f) Final illustrated image of sensor. (g) Cross-sectional view of sensor. (h) SEM image of the cross sectional view of sensor. (i) SEM image of the half middle part to show the second cavity at which the mechanical sensitive diaphragm touches first after pressure application. The bar scale of both the SEM images is 500 μm. Thickness of PI tape is found to be 15 μm, d = 425 μm, d+t = 451 μm because of the adhesiveness of PI tape. The thickness of single double sided tape is found to be 110 μm. The diaphragm is bent at the center due to pre-stress/build-in stress and gravity effect.

## IV. EXPERIMENTAL RESULTS AND DISCUSSION

A conventional scale from 0-25 was marked on the pneumatic pressure knob to mimic the structure of bourdon pressure gauge. A commercial MEMS pressure sensor (MS5803-14BA) was placed at the outlet air atomizing nozzle and then the marked scale from 0-25 was calibrated which corresponds to the applied pneumatic pressure. The commercially available MEMS capacitive pressure sensor was then replaced with our Al coated PI and paper sensor for characterization. The capacitance versus applied pressure response of sensor is shown in Fig. 3.

It can be observed from the results shown in Fig. 3 that the response of the sensor starts with nonlinear behaviour when the sensor was operating in normal region (shown in red) for the 0 – 7.5 kPa range. At a pressure of 7.5 kPa, the pull-in phenomena occurred, and the diaphragm touched the bottom electrode. This pressure is known as first touch pressure point (TP$_1$). At TP$_1$, the sensor entered the transition region in which the diaphragm touched the bottom cavity which is at the depth of $d + t$ (Fig. 1). As the pressure increased to 9.7 kPa, the diaphragm touched the PI tape which has a small concentric hole. This pressure is known as second touch pressure point (TP$_2$). The sensor then started to operate in the linear region from 14.24 kPa which is our range of interest for designing the integrated circuitry for any efficient and high precision application. Beyond the 54.9 kPa pressure point, the sensor operated in the saturation region because the maximum area of the mechanically sensitive diaphragm was already touching the bottom. In the linear region, the sensor has a sensitivity of 0.674 fF/Pa.

## V. WIRELESS BLUETOOTH-ENABLED CONTROL OF ROBOTIC ARM USING OUR PRESSURE SENSOR

In order to demonstrate the performance of our proposed polymer/paper-based double touch mode capacitive pressure sensor, it was integrated with a system to enable wireless control of a robotic arm. Normally, the actuators are used to achieve these kind of movements with micro-servomotors [41]. Servomotors are commonly used to precisely control the movement of robotic arms and joints. Our developed sensor can provide precise control of the movement of the servo-motors which enables the movement of the robotic arm. The robotic arm was built by assembling a tilt kit (Adafruit Industries). The motors were assembled within the arm in a way that allowed for two degrees of freedom (DOF) to the system. One servo enabled 180° side-to-side rotation and the other servo provided 150° upward or downward tilt angle. The latter angle was limited by the design of the tilt kit itself. A gripper was designed, 3D printed and mounted on the head to complete the assembly of the arm. In the transmitter unit, Arduino pro mini microcontroller was used to read capacitance values produced by the pressure sensor using Arduino capacitive sensing library. This change in capacitance was then mapped with the corresponding angle change of the servo position. A Bluetooth module (HC-05) was used to transmit this change obtained from the sensor to another Bluetooth module (HC-05) of the receiver unit in the form of a command to move the position of the servo according to the capacitance change. The two servomotors of the robotic arm were connected to the receiver unit which consisted of Arduino pro mini microcontroller, and a Bluetooth module. Each of the units was powered by a rechargeable lithium-ion battery (3.7V, 105mAh). The transmitter sensing unit is shown in Fig. 4 (a). Two pressure sensors were integrated with microcontroller, the Bluetooth module and the battery which were all placed at the bottom of the sensing element. The initial position of the robotic arm without applying any pressure is shown in Fig. 4(a). Upon applying pressure to sensor-1, the position of the servo that provided rotational movement changed. As seen in Fig. 4 (b) and after applying enough pressure, the position of the arm gripper changed to 180°. Upon applying pressure to sensor-2, however, the position of the servo that provides tilt movement changed to around 150° downwards. The linear range of pressure applied on sensor-1 and sensor-2 to actuate rotation and tilt movements in the robotic arm is shown in Fig. 5. The pneumatic pressure set-up used for the characterization of sensor was used to map the change in capacitance with the applied pressure. The microcontroller was used to measure and map the change in capacitance with the rotation and tilt angles of the motors which is shown in Fig 5.

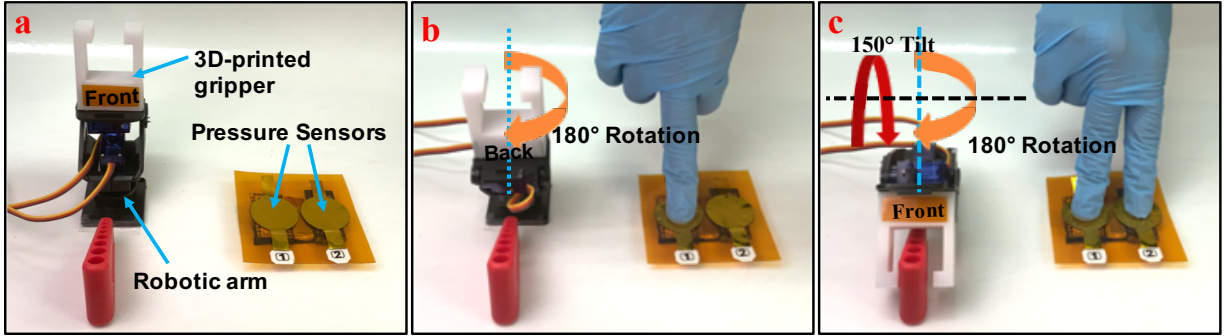

Fig. 4. Wireless control of robotic arm using double-touch mode capacitive pressure sensors. (a) Intial position of robotic arm with 3D printed gripper. Two pressure sensors are integrated with a microcontroller, Bluetooth module and battery to act as transmitter unit. (b) Rotational movement of robotic arm while applying pressure to sensor-1. (c) Rotational and tilt movement of robotic arm while applying pressure to sensor-1 and sensor-2 simultaneously. The gripper is now in close proximity to the object of interest after the applied movements.

## VI. Conclusion

The performance of double touch mode capacitive pressure sensor on pneumatic pressure application from (0 to 60 kPa) has been successfully demonstrated. Low-cost garage fabrication methods or DIY–electronics, which include cutting and taping of flexible substrates only, were utilized for fabrication of the sensor. The sensor was used as a wireless controller of robotic arm and joints. This presented application explains the potential of using proposed polymer/paper-based sensors in remote surgeries where precise control of robotic arms is required. Moreover, the sensor can be used in different devices and systems in medical, aerospace, and automobile industries due to its cost-effectiveness, high sensitivity and long linear range of performance. In the future, a third servomotor can be assembled into the 3D printed gripper in order to provide a third degree of freedom where the arm is able to control the grip to pick or drop certain desired objects. The performance and linear range of the sensor can be improved and optimized based on the desired application by scaling the larger diaphragm radius and inner circular radius of PI-tape, according to further application.

### Acknowledgment

Authors acknowledge Dr. Sherjeel M. Khan (currently Postdoctoral researcher at MMH Labs, KAUST) for helping out to design air-pressure experimental set-up.

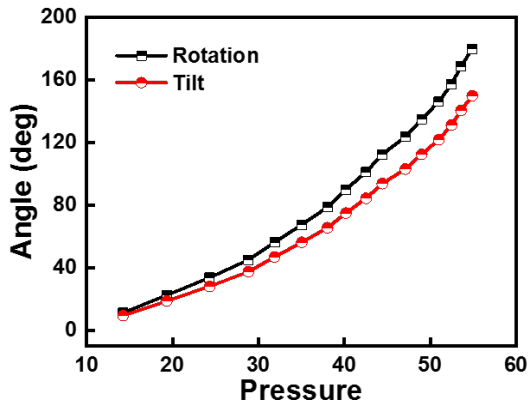

Fig. 5. Angle variation due to rotation and tilt of servomotors v/s applied pressure on the sensors in linear region i.e. when sensor operates after $2^{nd}$ touch pressure point (TP$_2$).